\documentclass[twocolumn]{aastex63}

\usepackage{graphicx}   
\hypersetup{linkcolor=red,citecolor=green,filecolor=cyan,urlcolor=magenta}

\newcommand{\co}{\mbox{$^{12}$CO}}
\newcommand{\coa}{\mbox{$^{13}$CO}}

\newcommand{\kms}{\mbox{km s$^{-1}$}}
\newcommand{\htwo}{\mbox{H$_2$}}

\newcommand{\cc}{\mbox{cm$^{-3}$}}

\newcommand{\msun}{\mbox{M$_\odot$}}

\received{}
\revised{}
\accepted{}
\submitjournal{ApJ}

\shorttitle{M51 Dense Gas}
\shortauthors{Heyer et al.}

\begin{document}

\title{The Dense Gas Mass Fraction and the Relationship to Star Formation in M51}

\correspondingauthor{Mark Heyer}
\email{heyer@astro.umass.edu}

\author[0000-0002-3871-010X]{Mark Heyer}
\affiliation{Astronomy Department, University of Massachusetts,
Amherst, MA, 01003 USA}
\author[0000-0003-4910-8939]{Benjamin Gregg}
\affiliation{Astronomy Department, University of Massachusetts,
Amherst, MA, 01003 USA}
\author[0000-0002-5189-8004]{Daniela Calzetti}
\affiliation{Astronomy Department, University of Massachusetts,
Amherst, MA, 01003 USA}
\author[0000-0002-1723-6330]{Bruce G. Elmegreen}
\affiliation{IBM Research Division, T.J. Watson Research Center, 1101 Kitchawan Road, Yorktown Heights, NY 10598, USA}
\author[0000-0001-5448-1821]{Robert Kennicutt}
\affiliation{Steward Observatory, University of Arizona, Tucson, AZ 85721-0065 USA}
\affiliation{George P. and Cynthia W. Mitchell Institute for Fundamental Physics \& Astronomy, 
Texas A\&M University, College Station, TX, 77843-4242, USA}
\author[0000-0002-8192-8091]{Angela Adamo}
\affiliation{The Oskar Klein Centre, Department of Astronomy, Stockholm University, AlbaNova, SE-10691 Stockholm, Sweden}
\author[0000-0003-2638-1334]{Aaron S. Evans}
\affiliation{Astronomy Department, University of Virginia, 530 McCormick Road, Charlottesville, VA, 22904, USA} 
\affiliation{National Radio Astronomy Observatory, 520 Edgemont Road, Charlottesville, VA, 22903, USA} 
\author[0000-0002-3247-5321]{Kathryn Grasha}
\affiliation{Research School of Astronomy and Astrophysics, Australian National University, Canberra ACT 2611, Australia}
\affiliation{ARC Centre of Excellence for All Sky Astrophysics in 3 Dimensions (ASTRO 3D), Australia}
\author[0000-0001-9969-3115]{James D. Lowenthal}
\affiliation{Department of Astronomy, Smith College, Northampton, MA 01063, USA}
\author[0000-0002-4723-6569]{Gopal Narayanan}
\affiliation{Astronomy Department, University of Massachusetts,
Amherst, MA, 01003 USA}
\author[0000-0003-1327-0838]{Daniel Rosa-Gonzalez}
\affiliation{Instituto Nacional de Astrof\'sica, \'Optica y Electr\'onica, Tonantzintla, 72840 , Puebla, Mexico} 
\author[0000-0002-2726-9443]{F.P. Schloerb}
\affiliation{Astronomy Department, University of Massachusetts,
Amherst, MA, 01003 USA}
\author[0000-0001-7915-5272]{Kamal Souccar}
\affiliation{Astronomy Department, University of Massachusetts,
Amherst, MA, 01003 USA}
\author{Yuping Tang}
\affiliation{Chinese Academy of Sciences South America Center for Astronomy, National Astronomical Observatories, CAS, Beijing, 100012, China}
\author[0000-0003-1774-3436]{Peter Teuben}
\affiliation{Department of Astronomy, University of Maryland,
College Park, MD, 20742, USA}
\author[0000-0002-2852-9737]{Olga Vega}
\affiliation{Instituto Nacional de Astrof\'sica, \'Optica y Electr\'onica, Tonantzintla, 72840 , Puebla, Mexico} 
\author[0000-0001-6142-397X]{William F. Wall}
\affiliation{Instituto Nacional de Astrof\'sica, \'Optica y Electr\'onica, Tonantzintla, 72840 , Puebla, Mexico} 
\author[0000-0001-7095-7543]{Min S. Yun}
\affiliation{Astronomy Department, University of Massachusetts,
Amherst, MA, 01003 USA}


\begin{abstract}
Observations of \co\ J=1-0 and HCN J=1-0 emission from NGC~5194 (M51) made with the 50~meter Large Millimeter Telescope and the 
SEQUOIA focal 
plane array are presented.  Using the HCN to CO ratio, we examine the dense gas mass fraction over a range of environmental conditions within the galaxy. 
Within the disk, the dense gas mass fraction varies along spiral arms but the average value over all spiral arms is comparable to 
the mean value of interarm regions.  We suggest that the near constant dense gas mass fraction 
throughout the disk arises from  
a population of density stratified, self gravitating molecular clouds and the required density threshold 
to detect each spectral line.
The measured dense gas fraction significantly increases in the central bulge in response to the 
effective pressure, $P_e$, from the weight of the stellar and 
gas components.
This pressure modifies the dynamical state of the molecular cloud population and possibly, the HCN emitting regions, in the central 
bulge from 
self-gravitating to diffuse configurations in which $P_e$ is greater than the gravitational energy density of individual clouds.
Diffuse molecular clouds comprise a significant fraction of the molecular 
gas mass in the central bulge, which may account for the measured 
sublinear relationships between the surface densities of the star formation 
rate and molecular and dense gas. 
\end{abstract}


\section{Introduction} \label{sec:intro}
The birth of a star is marked by the ignition of thermonuclear burning in the central core of a protostellar object.
This remarkable 
event is the endpoint 
of a sequence of processes that redistribute neutral interstellar 
gas into increasingly higher density configurations.  
Large, massive molecular clouds emerge from the neutral atomic component of the interstellar medium (ISM).
These molecular clouds fragment into 
higher density clumps and filaments \citep{Andre:2014}.  Within the clumps and filaments, high density pre-stellar 
cores develop 
that ultimately collapse under their own self-gravity to initiate the protostellar 
stage from which a star is ultimately produced \citep{Beuther:2007, Geiser:2021}.  
In each step, only a small fraction of the gas mass is converted into the 
next, higher density stage.   The low yields for these transitions
contribute to the measured inefficiency of star formation in which 
only $\sim$1\% of gas mass is converted into stars over a free-fall 
time scale as evaluated from whole galaxies to molecular clouds \citep{Krumholz:2012, Pokhrel:2021}.  

Observations of resolved molecular clouds in the solar neighborhood of the Milky Way 
have long demonstrated the spatial link between recent star formation and  localized pockets of
gas with volume densities greater than 10$^4$ \cc\ 
\citep{Myers:1983,
Bergin:2007, Wu:2010}. \citet{Gao:2004} extended this connection to galaxies by identifying a relationship between 
the infrared luminosity, a measure of the star formation rate, SFR, and the luminosity of HCN J=1-0 emission, 
a proxy for dense, molecular gas mass. 
Since that study, followup 
investigations with increasingly higher angular resolution, coverage, and sensitivity towards nearby galaxies have 
extended the relationship over 10 dex 
in HCN luminosity \citep{Jimenez-Donaire:2019}. 

Gas overdensities within molecular clouds are 
readily generated by the effects of gravity, the slow diffusion of the interstellar magnetic field, and 
supersonic, super-Alfv\'enic motions within a cloud. 
However, molecular clouds are not isolated objects but are part of the larger ecosystem of a galaxy. 
The local galactic environment can modulate the properties of molecular clouds \citep{Meidt:2016}.
Such environmental factors include the mid-plane pressure from the weight of stars, gas, and dark matter;  the action of spiral density 
waves in disk galaxies that compress and  redirect gas flows entering the spiral gravitational potential; and radiative and mechanical 
feedback from massive stars that 
inject momentum and energy into cloud volumes that drive and sustain turbulent flows. 

The galaxy NGC~5194 (M51a) is an appropriate target to study the role of the local environment in the development of over-dense regions in molecular clouds.  
M51a (hereafter, M51) exhibits a central bulge and  prominent spiral structure within a molecular gas-rich disk that is actively forming new stars 
\citep{Bigiel:2016}.  The tidal interaction of M51a with its companion, M51b (NGC~5195), is likely responsible for exciting the strong spiral 
density waves \citep{Dobbs:2010}. 
Its nearly face-on view provides a clear perspective of spiral arms and interarm regions.

In this contribution, we investigate the spatial distribution of the dense gas fraction derived from 
the HCN to CO ratio of luminosities 
within M51
using 
new data from the 50~meter Large Millimeter Telescope.  We investigate the role of spiral structure in modulating 
the formation and 
development of dense clumps and filaments over the ensemble of molecular clouds within our telescope beam 
 and the impact on the rate and efficiency of star formation. 
In \S2, we describe the data presented in this study.  
Images of \co\ and HCN J=1-0 emission are 
presented in \S3, along with descriptions of how the star formation rates and stellar mass surface densities are derived. 
Variations of the dense gas mass fraction with environment  are examined in \S4.
The scaling relationships between star formation and both molecular and dense gas are analyzed in \S5.  In \S6, we discuss 
our results.

\section{Data} \label{sec:data}
\subsection{Molecular Line Emission} \label{subsec:lmt}
Observations of \co\ J=1-0 and HCN J=1-0 emission from M51 were obtained with the 50 meter Large Millimeter Telescope 
(LMT) Alfonso Serrano 
between January and March 2020, using the 16 element focal plane array receiver SEQUOIA.
The half-power beam widths of the telescope at the line rest frequencies for CO (115.2712018~GHz) and HCN (88.630416~GHz) are 
12\arcsec\ and 16\arcsec\ respectively.  
The Wide-band Array Roach Enabled Spectrometer (WARES) was used to process the spectral information 
using the configuration with 800 MHz bandwidth 
and 391 kHz per spectral channel, which provides a velocity resolution of  1.1 \kms\ for CO and 1.3 \kms\ for HCN.  

For \co, ten maps covering the same 10\arcmin$\times$10\arcmin\ area were observed using 
On-the-Fly (OTF) mapping with all scanning along the R.A. axis. HCN imaging was comprised of multiple OTF 
maps covering three 7\arcmin$\times$4\arcmin\ overlapping areas centered on the nucleus of M51, 2\arcmin\ north of the nucleus, 
and 2\arcmin\ south of the nucleus.   Data were calibrated by a chopper wheel that allowed switching between the 
sky and an ambient temperature load.  The chopper wheel method introduces a fractional uncertainty of $\sim$10\% 
to the measured antenna temperatures \citep{Narayanan:2008}. 
Routine pointing and focus measurements 
were made to ensure positional 
accuracy and optimal gain. 

All data were processed with the LMT spectral line software package, which included zero-order baseline subtraction and 
the coadding of multiple maps into a final spectral line data cube at the native angular and spectral resolutions. 
To directly compare the \co\ and HCN intensities, we also processed the \co\ data to the HCN 
resolution $\lambda$/D=14\arcsec\ and $\lambda$/2D=7\arcsec\ sampling, where D is the 50~meter diameter of the LMT antenna.  
At a distance of 8.58$\pm$0.10~Mpc \citep{McQuinn:2016}, this resolution corresponds to a spatial size of 582~pc.
Both CO and HCN data cubes were spectrally smoothed and resampled to 5~\kms\ resolution. 
A main beam efficiency of 0.65 is applied to convert the data from 
T$_A^*$ temperature scale to main beam temperatures.   
The median rms sensitivities in main beam temperature units at 5~\kms\ spectral resolution are 28 milli-Kelvin (mK) 
for \co\ and 8~mK for HCN. 

\subsection{Ancillary  Data}
Our analysis requires several ancillary data sets.  To measure the unobscured star formation rates, we use far-ultraviolet (FUV) data taken from 
the GALEX Ultraviolet Atlas of Nearby Galaxies \citep{GildePaz:2007}.  The obscured star formation rate is derived from {\it Spitzer} MIPS 24\micron\ 
data from the Spitzer Infrared Nearby Galaxies Survey (SINGS) \citep{Kennicutt:2003}.  
Models of the galactic structure are derived from the 
3.6\micron\ image from SINGS. 
The mass of the stellar component of M51 is calculated using 
the {\it H} band image from 2MASS \citep{Jarrett:2003}, supplemented 
by {g, i} band images from the Sloan Digital Sky Survey (SDSS-III) 
\citep{Aihara:2011} and produced by \citep{Brown:2014}.  
All ancillary data sets were obtained from the 
NASA/IPAC Extragalactic Database (NED).

\section{Results} \label{sec:results}
\subsection{Molecular Line Images} 

\begin{figure*}
\centering
\includegraphics[width=0.75\textwidth]{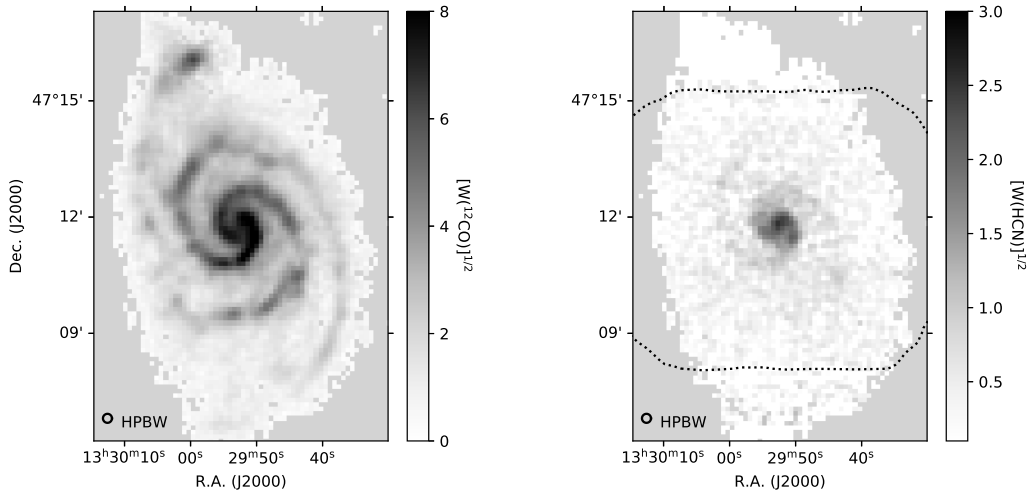}
\caption{(left) Masked image of (W(CO))$^{1/2}$. 
The square root 
of integrated intensity values is displayed in order to enhance the low surface brightness features.
The median uncertainty 
of WCO is 0.44~K km s$^{-1}$. 
(right) Masked image of (W(HCN))$^{1/2}$ using the \co-defined mask.  The median uncertainty of W(HCN) is 
0.10~K km s$^{-1}$. 
The dotted contour marks the 90$^{th}$ percentile of HCN 
rms values. 
\label{fig:fig1}}
\end{figure*}

Images of velocity-integrated CO J=1-0 emission and HCN J=1-0 emission are displayed in Figure~\ref{fig:fig1}.  These images are 
created by masking the spectral channels where CO is detected for each spectrum in the coadded map in order to limit contributions 
of noise that can occur when integrating over a fixed, wide velocity interval of a rotating galaxy.
This masking is achieved by the following steps described by \citet{Dame:2011}.   First, the CO data are smoothed along the angular 
axes using a Gaussian kernel with a half power beam width (HPBW) of 5 pixels (35\arcsec) and along the spectral axis using a boxcar 
kernel with a width of 3 spectral channels (15 \kms) while keeping the angular and spectral sampling of 7\arcsec\ and 5~\kms.  
For a given spectrum in the smoothed data cube,  any grouping of 3 or more consecutive channels 
with brightness temperatures greater than 3.75$\sigma$, 
are flagged as active voxels 
and assigned a value of 1 in the mask data cube, $M(x,y,v)$.  All other channels are assigned a value of zero. 
This signal to noise threshold is a result of many trials with varying signal to noise ratios to maximize the number of active pixels while minimizing false detections. 
The integrated intensity image, $W$, is calculated from the data cube, $T_{mb}(x,y,v)$ from the expression, 
\begin{equation}
W(x,y) = {\Delta}v\displaystyle\sum_{k} T_{mb}(x,y,v_k)M(x,y,v_k), 
\end{equation}
where the sum is over all spectral channels
and 
${\Delta}v$ is the 
channel width of 5~\kms.  The statistical error of this integrated intensity is $\sigma(x,y) {\Delta}v (N_c(x,y))^{1/2} $
where $\sigma(x,y)$ is the root mean square of channels outside the velocity range 360-800~\kms\ of the unsmoothed data cube,
 and $N_c(x,y)$ is the number of active channels
for position x,y.  

The left panel in Figure~\ref{fig:fig1} shows widespread detection of \co\ emission over the projected area of M51.  The brightest 
emission 
traces the bulge and inner spiral features identified in previous studies with comparable 
resolution  \citep{Koda:2011, Pety:2013}.  Signal is also detected within the interarm regions -- even within the 
large, interarm area in the south.  Integrating the CO emission over all pixels, 
we derive a CO luminosity of (2.167$\pm$0.003)$\times$10$^9$ Kkms$^{-1}$pc$^2$, assuming a distance of 8.58~Mpc. 
The quoted uncertainty is based on propagating observational noise rms values through the luminosity calculation.
This total CO luminosity includes contributions from NGC~5195. 
The total molecular mass is  
(9.32$\pm$0.01$)\times$10$^9$ \msun\ using the \htwo-to-CO conversion factor $\alpha_{CO}$=4.3 \msun/(Kkms$^{-1}$pc$^2$) \citep{Bolatto:2013}.
The CO luminosity of just NGC~5195 is (9.93$\pm$0.01)$\times$10$^7$ Kkms$^{-1}$pc$^2$, which corresponds to a molecular 
mass of (4.27$\pm$0.04$)\times$10$^8$ \msun.
The CARMA-NRO CO imaging estimated the total mass of NGC~5194 and NGC~5195 to be 6.0$\times$10$^9$~\msun\ 
applying the same CO-\htwo\ conversion factor and distance as used in this study \citep{Koda:2011}.
Within calibration uncertainties, masking thresholds, and map coverage, our values are also comparable with 
\citet{Pety:2013}, who derived L$_{CO}$=1.8$\times$10$^9$ Kkms$^{-1}$pc$^2$, 
and molecular mass of 7.9$\times$10$^9$ \msun\
from observations with the IRAM 30~m 
telescope that covered approximately the same area.

The \co-defined mask is applied to the HCN data from which we produce a velocity-integrated image of HCN emission shown in 
the right panel of Figure~\ref{fig:fig1}.  For areas with declinations south of 47$^\circ$~07\arcmin\ and 
north of 47$^\circ$~15\arcmin\  
that includes NGC~5195, 
the HCN noise levels are large due to less data collected near these map edges.
These areas are not included in any subsequent analyses presented here. 
The HCN emission is strong in the central region of M51 but is more patchy and uneven with increasing galactic radius than the CO emission 
yet still exhibits faint spiral structure in the 
central 3\arcmin$\times$3\arcmin\ area.
The HCN luminosity over the map area is (6.05$\pm$0.08)$\times$10$^7$ Kkms$^{-1}$pc$^2$.  Adopting the conversion 
factor, $\alpha_{HCN}$=10 \msun/(K km s$^{-1}$ pc$^2$) that assumes self-gravitating dense cores in clouds 
\citep{Gao:2004a}, 
the dense gas mass is (6.05$\pm$0.08)$\times$10$^8$ \msun. 

\subsection{Star Formation Rates} 
Star formation rates (\msun/yr) within M51 are derived from images in the far ultraviolet (FUV) band
measured by {\it GALEX} and MIPS 24\micron\ band from the {\it Spitzer Space Telescope}.  
In regions of low dust obscuration, massive young  stars 
directly contribute to the observed FUV flux.  Newborn 
stars embedded within 
regions of moderate to high extinction contribute to
the 24\micron\ flux as their stellar optical and UV radiation heat nearby dust grains that reradiate this energy into  
the mid and far infrared bands.  
Used together, these bands account for star formation rates from both unobscured and deeply embedded regions of 
star formation. 
However, an older stellar population that does not reflect star formation over the last 100~Myr also contributes
to the measured luminosities in both the FUV and 24\micron\ bands. 
To derive more accurate star formation rates, it is necessary to account for this ``cirrus'' emission component from the 
old stellar population in each band.

To evaluate this contribution, we construct 3 models of the star formation rate.
Model 0 is the trivial model which assumes no contributions to the FUV and 24\micron\ luminosities from older stars.  
This model allows a comparison to other studies that also exclude this component \citep{Bigiel:2016, Jimenez-Donaire:2019}. 
The other models consider the distribution of the old, stellar component by decomposing the 
{\it Spitzer} image of 3.6\micron\ surface brightness into 
galactic structural components 
using {\tt Galfit} \citep{Peng:2010}. 
Model 1 follows the example in \citet{Peng:2010} for M51 that 
includes a 2 component central bulge and inner and outer spiral arms.
We exclude the numerous, bright, 3.6\micron\ emission knots so that 
each {\tt Galfit} component more accurately 
tracks the smooth emission, which limits oversubtractions.
Model 2 includes a 2 component central bulge and a Sersic profile that approximates an 
exponential disk but no spiral arm structures.  
In this model, we implicitly assume that the spiral arms are entirely due to young stellar populations, 
although this assumption is an over-simplification of current results for spiral structures \citep{Kreckel:2016}. 
In calculating the exponential disk,
we exclude pixels located 
in spiral arm features (see Figure~\ref{fig:fig3}) so as 
to not bias the fit.   
For both {\tt Galfit} models, we also simultaneously fit the companion galaxy,  M51b,  with a 2 component central bulge, 
and one spiral arm component, based on the \citet{Peng:2010} models.  This avoids biasing the M51a fits in the outer disk and spiral 
arm components. 
The {\tt Galfit} parameters and 
the 3.6\micron\ model images for Models 1 and 2 are shown in Appendix~\ref{appendixA}.

The 3.6\micron\ {\tt Galfit} model intensities are scaled to the FUV and 24\micron\ bands.  The respective scaling factors 
are determined from the ratio of the median observed FUV and 24\micron\ intensities to the median model 3.6\micron\ 
intensity within several 5$\times$5 pixel areas 
located within interarm regions with no star formation activity.
The resultant scaled models 
are then subtracted from the observed images to remove the old stellar population contributions.
For this subtraction, we include only model components where we expect a significant contribution to the FUV and 24\micron\ 
intensities from the old stellar populations. For 24\micron,  we subtract the normalized bulge and both spiral components 
for Model 1 and the bulge and disk for Model 2.  
It is clear from the lack of a significant bulge in the FUV image that the unobscured, central stellar population is old and is 
contributing little UV flux. So for the FUV, we subtract only the two spiral components for Model 1 and the disk for Model 2. 
Since Model 2 does not account for spiral arm features in which older stellar populations are enhanced, 
its subtraction leads to a positive bias of SFR values.
All pixels with negative residuals are set to 
zero to limit artificial model bias in the SFR maps resulting from the image combination. 
Using the \citet{Aniano:2011} kernels, the corrected 24\micron\ and FUV images are convolved to the resolution 
(14\arcsec) and pixel 
size (7\arcsec) of the molecular line data.

For each model,
the star formation rate is calculated 
using the expression by \citet{Liu:2011}, 
\begin{equation}
SFR(M_\odot yr^{-1}) = 4.6\times10^{-44}[L(FUV)+6.0L(24{\mu}m])
\end{equation}
where the luminosities are in ergs-s$^{-1}$. This calibration
assumes a \citet{Kroupa:2001} stellar IMF to account for newly formed low mass stars 
that do not significantly contribute to the FUV and 24\micron\ emission.  
The uncertainty of SFR is 1.33$\times$10$^{-5}$ M$_\odot$ yr$^{-1}$, which is derived from the 
standard deviation of the 
background areas in both the FUV and 
24\micron\ images and propagated through equation 2.  

We acknowledge that our method to quantify the old stellar population contributions to the star formation tracers is 
complicated and depends on the accuracy of the decomposition of the 3.6\micron\ image by {\tt galfit} as well as the scaling 
of the model 3.6\micron\ surface brightness values to the FUV and 24\micron\ bands. Alternatives to account for the cirrus 
emission include unsharp masking to identify a diffuse component produced by radiation from the old stellar population  
\citep{Rahman:2011} 
and direct modeling 
of the cirrus emission from radiation fields and dust emissivities 
\citep{Leroy:2012}.
Our method 
offers a more physically-based alternative to 
unsharp masking 
as {\tt galfit} identifies galactic components that are expected to be populated by 
older stars. 

\subsection{Stellar Mass Surface Density} 
The mass of the stellar component is derived using the method described by 
\citet{Zibetti:2009}.  In brief, {\it g, i, H} band images are resampled to the 3.6\micron\  
image pixel size and coverage.  Background and foreground objects identified in the 3.6\micron\ image are 
removed in the {g, i, H} images by interpolation.  The $g$ and $i$ images are convolved to the 
2.8\arcsec\ resolution 
of the {\it H} band data and all images are converted to Jy/pixel surface brightness units. 
Each image is adaptively smoothed
to 
achieve signal to noise ratios greater than 20 using the code of \citet{Zibetti:2009b}.  For a given pixel, the largest smoothing element is 
selected and applied to the remaining images in order to match the spatial resolution.  In all cases across 
the disk of M51, the 
{\it H} band data required the largest smoothing kernel. 
The $g-i$ and $i-H$ colors are derived for each pixel and are corrected for galactic foreground
extinction assuming A$_v$=0.096 \citep{Schlafly:2011}.  From the colors, the median mass to light ratio 
at H band, $\Psi_H$,  
is determined based on the look-up tables of \citet{Zibetti:2009} that are based on the \citet{Bruzual:2003} 
stellar population synthesis models.
For pixels with colors outside 
the look-up table limits, values for $\Psi_H$ are interpolated from valid, neighboring pixel values. 
{\it H} band luminosities, $L_H$,  are calculated from the adaptively smoothed H band image 
using a distance of 8.58~Mpc. 
We convert these luminosities to ``in-band'' solar units using L$_{\odot,H}$ = 1.08$\times$10$^{33}$ ergs-s$^{-1}$,
 derived from the 
absolute Vega magnitude of the Sun in the 2MASS H band and the flux zero-point listed in \citet{Willmer:2018}. 
The stellar mass image is $L_H \Psi_H$.  The resultant stellar mass 
image is then smoothed and resampled to the pixel size and resolution of the molecular line data. 

The primary advantage of using {\it H} band to derive stellar masses is the method does not require a 
subtraction 
of an extended, warm dust component that is heated by star formation activity as is the case at 3.6\micron\
\citep{Meidt:2012, Querejeta:2015}, which can introduce additional 
uncertainties in the final result. However, owing to low signal to noise of the 2MASS {\it H} band data in 
the outer regions of the disk, this band is not as useful as the highly sensitive 3.6\micron\ image 
for decomposing the galaxy into structural components using 
{\tt Galfit} as described in \S3.2.

\section{Dense Gas Mass Fraction in the Molecular Cloud Population of M51}
The distribution of high 
density ($>$10$^4$ \cc) regions within molecular clouds provides a roadmap for both ongoing star formation 
and star formation activity to occur within several free-fall times of the dense gas.
Integrated over the projected area of a cloud or an ensemble of clouds, the dense gas mass fraction can offer a measure of 
future star formation efficiency assuming other conditions are satisfied such as self-gravitating clumps and cores and a 
fixed fraction of dense gas redistributed into stars. 
We define the dense gas fraction as the 
ratio of cloud mass residing in gas configurations with number densities greater than $n^\prime$ to the total cloud mass, 
\begin{equation}
f_{DG}(n>n^\prime)=M(n>n^\prime)/M_{TOT}
\end{equation}
where $n^\prime$ is a threshold volume density appropriate for the dense gas tracer.  
For a single, spectral line high density gas tracer such as the HCN J=1-0 line used in this study,  
this expression assumes a small contribution to the dense gas mass from regions with densities much larger than the 
critical density 
required to collisionally excite the molecules into the upper energy level of the transition. 
The measured $f_{DG}$ value represents a luminosity-weighted average of the dense gas mass fraction for the 
set of molecular clouds within the telescope beam or over the area subtended by the stacking condition (see \S4.1). 

The HCN to CO intensity ratio has been widely applied as a measure of the dense gas fraction in 
Galactic molecular clouds
\citep{Jackson:1996, Helfer:1997, Paglione:1998, Evans:2020}.  
The CO luminosity 
provides an estimate of the bulk mass of molecular clouds while the HCN J=1-0 line (or higher rotational transitions) 
tracks the mass of the higher volume density component within a cloud.  
Specifically, 
\begin{equation}
f_{DG}(n>n^\prime) = \frac{M_{dense}}{M_{mol}}=\frac{\alpha_{HCN} L_{HCN}}{\alpha_{CO} L_{CO}} = (2.3\pm1.3) R
\end{equation}
where $R$ is the observed HCN to CO ratio of luminosities and the \htwo\ component with low 
CO abundances in diffuse gas is not included.   The coefficient and uncertainty terms in equation 4 assume 
$\alpha_{CO}$=4.3 \msun/(K km s$^{-1}$ pc$^2$), $\alpha_{HCN}$=10 \msun/(K km s$^{-1}$ pc$^2$) and fractional 
uncertainties of 0.3 \citep{Bolatto:2013} and 0.5 respectively.
This optimistic conversion error still exceeds any 
measurement errors to derive $R$ and so the fractional uncertainties of $f_{DG}$ are greater than $>$75\%. 
We apply this factor for all conversions of $R$ to $f_{DG}$ in \S4.1.1, \S4.1.2, \S5.3, and \S6.

The appropriate volume density threshold, $n^\prime$, for the 
HCN J=1-0 line depends on several factors.
The critical density, $n_c$, of an emission line is 
the density at which the rate of collisional excitations 
equals the spontaneous decay rate.
For the HCN J=1-0 line, $n_c$=5$\times$10$^5$~\cc\ in the 
limit of optically thin emission and collisions with \htwo\ molecules and temperature 10-20~K \citep{Shirley:2015}. 
This density is comparable to the mean density of 
pre-stellar and protostellar cores, 
which makes this line so attractive as a dense gas tracer.
However, observations of the HCN J=1-0 line in the Milky Way demonstrate that the line is often optically thick so 
one must account for radiative trapping and its impact on line excitation to derive an effective critical density. 
For the HCN line, this 
effective critical density is 
$\sim$(3-10)$\times$10$^3$~\cc\ for kinetic temperatures less than 20~K, varying column densities and HCN abundances, and collisions 
with \htwo\ molecules 
\citep{Helfer:1997, Paglione:1998, Shirley:2015}.  
These densities are more similar to the dense clumps and filaments within molecular clouds rather than the pre-stellar or protostellar 
cores.  
In addition, \citet{Goldsmith:2017} demonstrate that in low density environments 
with high ionization fractions such as envelopes of molecular clouds or in the central regions of galaxies, 
electron collisions can also impact the HCN excitation, which can further decrease the effective critical density of 
the HCN J=1-0 line.  \citet{Goicoechea:2021} find similar results when electron collisions dominate the 
excitation of the HCN molecules.  
For this study, we adopt the value of $n^\prime=10^4$~\cc\ for the HCN J=1-0 transition that assumes a mixture of optically 
thin and thick emission and no electron excitation.  For the optically thick CO emission, radiative trapping 
maintains sufficient excitation such that the effective critical density is $\sim$100 \cc. 

\begin{figure}
\centering
\includegraphics[width=0.4\textwidth]{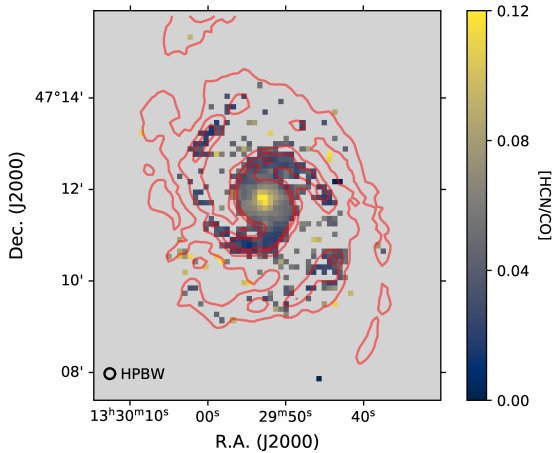}
\caption{Image of the HCN/CO ratio at 14\arcsec\ angular resolution. Only pixels with HCN to CO ratios greater than 3$\sigma$ are 
shown.  Red contours show the \co\ J=1-0 
integrated intensities to delineate spiral arm features.  The contour levels are 5,15,25,35 K-\kms.
\label{fig:fig2}}
\end{figure}

To examine the dense gas fraction in M51, we calculate 
$R$ for each 7\arcsec\ pixel in the matched HCN and CO images of integrated intensity and propagate intensity uncertainties to derive 
$\sigma(R)$.
The sensitivity of $R$ is limited by the much weaker HCN emission. 
The image of $R$ is shown in Figure~\ref{fig:fig2}, where we have clipped pixels with 
$R/\sigma(R)$ 
less than 3. 
The red contours show the distribution of $W(^{12}CO)$ that traces the spiral structure. 
The highest ratios ($R>$0.08) are found in the central 
bulge area of M51.  Beyond the extent of the bulge, spatially coherent features of $R$ follow the 
inner spiral arms.  
There are marginal 
detections of $R$ in a few locations between the spiral arms.  

\subsection{Spectral Stacking}
The HCN to CO ratio can be probed to fainter levels of the HCN emission than the 3$\sigma$ limits of $R$ shown in Figure~\ref{fig:fig2} 
by stacking the HCN and \co\ spectra selected by a common physical attribute or spatial grouping. 
The stacked spectra represent the average of the ensemble 
that satisfy the selection criteria. 
This spectral stacking method has been applied to HCN data by \citet{Bigiel:2016} and \citet{Jimenez-Donaire:2019}, where HCN 
spectra are stacked in bins of galactic radius or stellar surface density.  

\subsubsection{Bulge and Spiral Arm Regions of Interest}
\begin{figure*}
\centering
\includegraphics[width=0.75\textwidth]{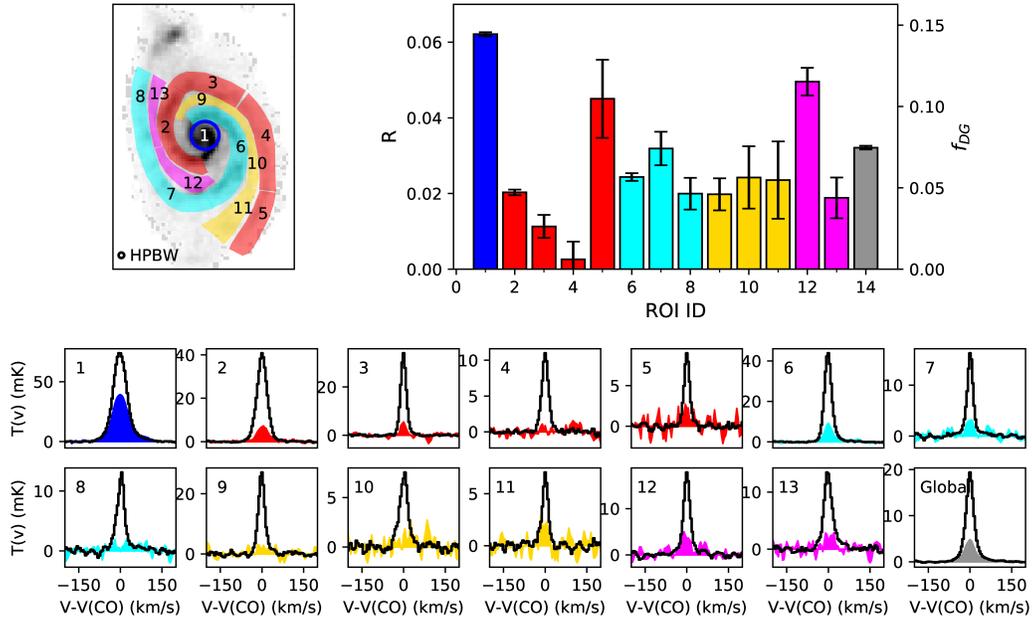}
\caption{(top left) Image of \co\ J=1-0 integrated intensity and regions of interest 
corresponding  to the central bulge (blue circle), spiral arm segments (red and cyan shading), 
and interarm segments (yellow and magenta shading). 
(Bottom) Stacked \co\ (black and divided by 10) and HCN spectra (colored according to group) 
for each ROI.  The global CO and HCN spectra 
correspond to the stacked CO and HCN spectra (grey) for all 
pixels with detected \co\ emission.
(top right) Ratio of HCN to CO luminosities (left axis) and $f_{DG}$ (right axis) derived from stacking CO and HCN emission in 
each ROI. Error bars refer to 
uncertainties in R values.  The grey bar (ROI 14) 
is the global HCN/CO value. 
\label{fig:fig3}}
\end{figure*}

The deep gravitational potential of spiral arms provides an environment that could impact the dense gas fractions of the 
molecular cloud population.  To explore this role, we define regions of interest (ROI) in the 
3.6\micron\ image as polygons that 
circumscribe 
segments along both spiral arms and interarm regions. 
We also define 
the M51 bulge as a distinct ROI that includes the area within a circular radius of 35\arcsec.  
Finally, all CO and HCN spectra within the CO-defined mask (5997 pixels) are stacked to produce globally averaged
 spectra (ROI~14).
The top left panel 
of Figure~\ref{fig:fig3} shows the labels for all the ROIs, which are also grouped 
into separate arms and interarms distinguished by color. 

For each ROI, we identify the set of pixels that fall within the subtended area.  
In the case where the distance between the pixel position and the nearest ROI 
boundary is less than 7\arcsec\ corresponding to half of the half-power beam width,
we do not correct for signal gathered by the 
telescope beam that extends beyond the ROI boundary.
To stack the data, the velocity 
axis for each selected spectrum in the ROI is shifted to be centered at 0 \kms\ by subtracting the velocity centroid of the CO 
spectrum at the same position. This relative spectrum is sampled between -200 and +200 \kms\ 
with 5~\kms\ spectral resolution and coadded to the stacked spectrum with 1/$\sigma^2$ weighting.
The bottom subplot in Figure~\ref{fig:fig3} shows the stacked \co\ and HCN spectra.
The HCN spectra are colored according to the ROI groupings. 

The resultant stacked \co\ and HCN spectra are integrated over velocities where CO is detected in the stacked spectrum.
The rms values of the stacked spectra are derived from the standard deviation of intensities outside of these velocity intervals. 
The HCN to CO ratio is calculated from the ratio of HCN to CO integrated intensities.  
Table~\ref{tab:table1} summarizes the selected velocity intervals, measured intensities, ratios, and uncertainties for each ROI.

\begin{table*}[hbt]
\centering
\caption{ROI Stacked Spectra Properties \label{tab:table1}}
\begin{tabular}{rrccccccccc}
\hline
\hline
ROI & Npix & [v1,v2] & W(HCN) & $\sigma$(W(HCN)) & W(CO) & $\sigma$(W(CO)) & R & $\sigma$(R) & $f_{DG}$ & $\sigma(f_{DG})$ \\
 & & (\kms) & (K\kms) & (K\kms) & (K\kms) & (K\kms) & &  & &\\
\hline
 1 & 69 & [-125, 125] & 2.982 & 0.025 & 48.014 & 0.102 &  0.062 &  0.001 &  0.144 &   0.084\\ 
 2 & 188 & [ -90,  90] & 0.376 & 0.014 & 18.500 & 0.070 &  0.020 &  0.001 &  0.047 &   0.028\\ 
 3 & 158 & [ -75,  75] & 0.100 & 0.027 & 8.803 & 0.074 &  0.011 &  0.003 &  0.026 &   0.015\\ 
 4 & 199 & [ -50,  50] & 0.009 & 0.016 & 3.357 & 0.056 &  0.003 &  0.005 &  0.006 &   0.004\\ 
 5 & 144 & [ -50,  50] & 0.104 & 0.024 & 2.306 & 0.062 &  0.045 &  0.010 &  0.105 &   0.061\\ 
 6 & 262 & [ -75,  75] & 0.366 & 0.015 & 15.018 & 0.064 &  0.024 &  0.001 &  0.057 &   0.033\\ 
 7 & 215 & [ -70,  70] & 0.155 & 0.021 & 4.847 & 0.083 &  0.032 &  0.004 &  0.074 &   0.043\\ 
 8 & 105 & [ -70,  70] & 0.094 & 0.020 & 4.704 & 0.110 &  0.020 &  0.004 &  0.046 &   0.027\\ 
 9 & 61 & [ -60,  60] & 0.160 & 0.034 & 8.083 & 0.128 &  0.020 &  0.004 &  0.046 &   0.027\\ 
10 & 67 & [ -65,  65] & 0.068 & 0.023 & 2.796 & 0.087 &  0.024 &  0.008 &  0.056 &   0.033\\ 
11 & 139 & [ -60,  60] & 0.045 & 0.019 & 1.897 & 0.077 &  0.024 &  0.010 &  0.055 &   0.032\\ 
12 & 96 & [ -75,  75] & 0.293 & 0.021 & 5.911 & 0.118 &  0.050 &  0.004 &  0.115 &   0.067\\ 
13 & 69 & [ -80,  80] & 0.102 & 0.029 & 5.420 & 0.104 &  0.019 &  0.005 &  0.044 &   0.026\\ 
14 & 5997 & [-125, 125] & 0.049 & 0.001 & 1.522 & 0.007 &  0.032 &  0.000 &  0.075 &   0.044\\ 
\hline
\end{tabular}
\end{table*}

\co\ emission 
is readily detected in all ROIs that span most of the bulge and disk areas of M51 -- including 
all of the interarm regions.  The ubiquity of CO in the stacked spectra is expected given that all of the ROI areas fall within the 
CO-defined mask described in \S3.1.  From
extinction features within high resolution, optical images obtained with the Hubble Space Telescope ({\it HST}) and 
8\micron\ emission features from {\it Spitzer},
the interarm regions of M51 contain narrow spiral spurs of material that stretch 
between adjacent spiral arm segments.
These molecular features are not resolved by the \co\ data presented in this study 
but are evident in the Plateau de Bure Interferometer Arcsecond Whirlpool 
Survey (PAWS) \co\ image of M51 \citep{Schinnerer:2013}.  
Such spiral spurs emerge from the gas subjected to increased velocity shear in the 
interarm regions \citep{Kim:2002}. 
In M51, molecular clouds residing within these spurs are not destroyed by the 
high velocity shear of the interarm regions but maintain sufficient column densities for self-shielding to remain in the molecular gas phase throughout 
their transit through the interarm region to the next spiral arm 
\citep{Koda:2009}.  HCN emission is detected with signal to noise greater than 3 within most ROIs.  It is 
not detected at this level in ROIs 4,10, and 11. 

The top right subplot of Figure~\ref{fig:fig3} displays the $R$ values (left axis) and $f_{DG}$ values (right axis).
The error bars correspond to uncertainties in $R$. 
The large values of $R$ and $f_{DG}$ in the M51 bulge has been previously established in earlier studies 
\citep{Bigiel:2016, Gallagher:2018, Jimenez-Donaire:2019, Querejeta:2019}.
Within the uncertainites, there is little variation of $f_{DG}$ between spiral arms and interarms.
The dense gas fraction averaged over all spiral arm and interarm segments are 0.051
and 0.063 respectively. 
The comparable values of $f_{DG}$ is unexpected 
given the deeper gravitational potential that defines the spiral arms.
We discuss this limited range in $f_{DG}$ values in more detail in \S6.2.

\subsubsection{Stellar Mass Surface Density}
The scale height of the molecular ISM in galaxies is small in comparison to the distribution of stars and atomic gas.
Consequently, molecular clouds are subject to the effective pressure generated by the weight of the 
stellar, gas, and dark matter components residing above and below the disk mid-plane.  Such pressure 
facilitates the transition of warm, neutral, atomic gas into the cold, neutral atomic gas phase, which is often a 
precursor to the formation of molecular clouds \citep{Elmegreen:1993}. Can such pressure also impact the development 
of high density gas within molecular clouds? 

\begin{figure*}
\centering
\includegraphics[width=0.75\textwidth]{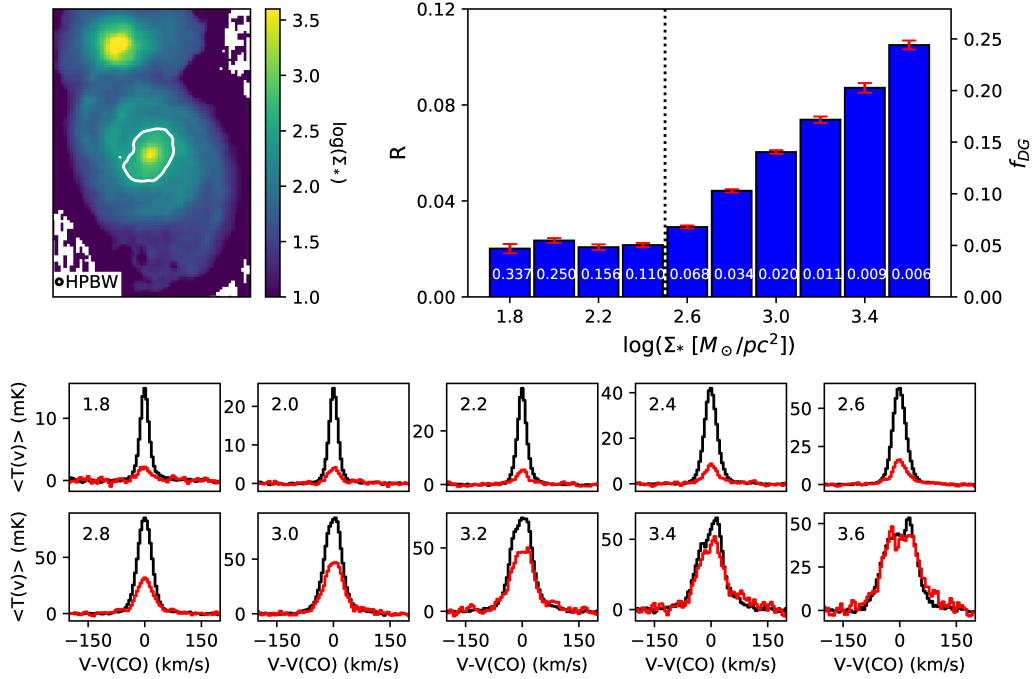}
\caption{(top left) Image of stellar mass surface density derived from the {\it 2MASS H} band image and a varying mass to light ratio 
based on $g-i$ and $i-H$ colors \citep{Zibetti:2009}.  The white contour denotes log($\Sigma_*$)=2.5.
(top right) HCN to CO luminosity ratio (left axis) and the dense gas fraction, $f_{DG}$ (right axis) in each bin of 
stellar mass surface density. Values in each bin reflect the 
fractional area covered.  The vertical dotted line marks the transition at log($\Sigma_*$) 
to higher $f_{DG}$ values with increasing stellar mass in the 
central bulge.
(Bottom) Stacked \co\ (black and divided by 10) and HCN spectra (red) 
for each stellar mass bin.  The number in each box is log($\Sigma_*$) for that bin.
\label{fig:fig4}}
\end{figure*}

Previous studies of HCN and CO 
in nearby galaxies have already demonstrated a relationship between $f_{DG}$ and the mass surface density of stars as well 
as the full pressure component, $P\propto\Sigma_{gas}\Sigma_*$ \citep{Gallagher:2018, Jimenez-Donaire:2019}.  
Here, we examine the relationship of the dense gas fraction with stellar mass surface density using spectral stacking.
Figure~\ref{fig:fig4} shows the stacked HCN and CO spectra 
for each log($\Sigma_*$) bin.  Table~\ref{tab:table2} summarizes the integrated intensities and ratios derived from 
the stacked spectra.
Within the disk, 
where log ($\Sigma_*$) $<$2.5 that contains 85\% of the area over all bins, $f_{DG}$ is $\sim$0.05 with little variation
over the log($\Sigma_*$) range of 1.8 to 2.4.  For log($\Sigma_*) >$ 2.5, corresponding to much of the stellar bulge 
component, the dense gas mass fraction 
rapidly rises.
Evaluating the dense gas fraction within 1~kpc wide bins of radius shows a 
similar constant value of $\sim$0.05 for radii between 2 and 8~kpc and a steep rise to 0.18 in 
the central 2~kpc. 
Both \citet{Bigiel:2016} and \citet{Jimenez-Donaire:2019} find similar profiles of $f_{DG}$ with galactic 
radius. 

\begin{table*}[hbt]
\centering
\caption{Stellar Mass Surface Density Stacked Spectra Properties  \label{tab:table2}}
\begin{tabular}{rrccccccccc}
\hline
\hline
log($\frac{\Sigma_*}{{\msun}pc^{-2}}$) & Npix & [v1,v2] & W(HCN) & $\sigma$(W(HCN)) & W(CO) & $\sigma$(W(CO)) & R & $\sigma$(R) & $f_{DG}$ & $\sigma(f_{DG})$ \\
 & & (\kms) & (K\kms) & (K\kms) & (K\kms) & (K\kms) & & & & \\
\hline
1.8 & 891 & [ -80,80] & 0.096 & 0.009 & 4.794 & 0.051 &  0.020 &  0.002 &  0.047 &   0.027\\ 
2.0 & 661 & [ -80,80] & 0.184 & 0.008 & 7.805 & 0.050 &  0.024 &  0.001 &  0.055 &   0.032\\ 
2.2 & 413 & [ -80,80] & 0.244 & 0.014 & 11.742 & 0.046 &  0.021 &  0.001 &  0.048 &   0.028\\ 
2.4 & 290 & [ -80,80] & 0.398 & 0.016 & 18.408 & 0.068 &  0.022 &  0.001 &  0.050 &   0.029\\ 
2.6 & 179 & [ -90,90] & 0.860 & 0.019 & 29.466 & 0.087 &  0.029 &  0.001 &  0.068 &   0.040\\ 
2.8 & 91 & [-100,100] & 1.890 & 0.029 & 42.737 & 0.158 &  0.044 &  0.001 &  0.103 &   0.060\\ 
3.0 & 52 & [-105,105] & 3.249 & 0.039 & 53.789 & 0.216 &  0.060 &  0.001 &  0.140 &   0.082\\ 
3.2 & 28 & [-140,140] & 4.091 & 0.074 & 55.390 & 0.212 &  0.074 &  0.001 &  0.172 &   0.100\\ 
3.4 & 23 & [-140,140] & 4.427 & 0.096 & 50.800 & 0.397 &  0.087 &  0.002 &  0.203 &   0.118\\ 
3.6 & 16 & [-150,150] & 5.380 & 0.091 & 51.246 & 0.285 &  0.105 &  0.002 &  0.244 &   0.142\\ 
\hline
\end{tabular}
\end{table*}

\section{Star Formation Scaling Relationships}

The Kennicutt-Schmidt (KS) scaling relationship connects the star formation rate with the amount of neutral gas mass (atomic, molecular, or 
both)
\citep{Schmidt:1959, Kennicutt:1989, Kennicutt:1998}.
It has been applied to disk-averaged values of $\Sigma_{SFR}$ and $\Sigma_{gas}$  for a large 
set of galaxies \citep{Kennicutt:1989,delosreyes:2019} and
resolved measures of $\Sigma_{SFR}$ and $\Sigma_{gas}$ values {\it within} a galaxy \citep{Kennicutt:2007, Leroy:2008, Bigiel:2008}.
The KS relationship is expressed as a power law, $\Sigma_{SFR}=A (\Sigma_{gas})^N$, whose parameters, $A$ and $N$,
and the measured scatter  provide
important constraints to the evolution of the ISM and pathways to star formation within the varying gas environments of galaxies 
\citep{Kennicutt:2012}.

\subsection{$\Sigma_{SFR}$ and $\Sigma_{mol}$}
\begin{figure*}[tbh]
\centering
\includegraphics[width=0.75\textwidth]{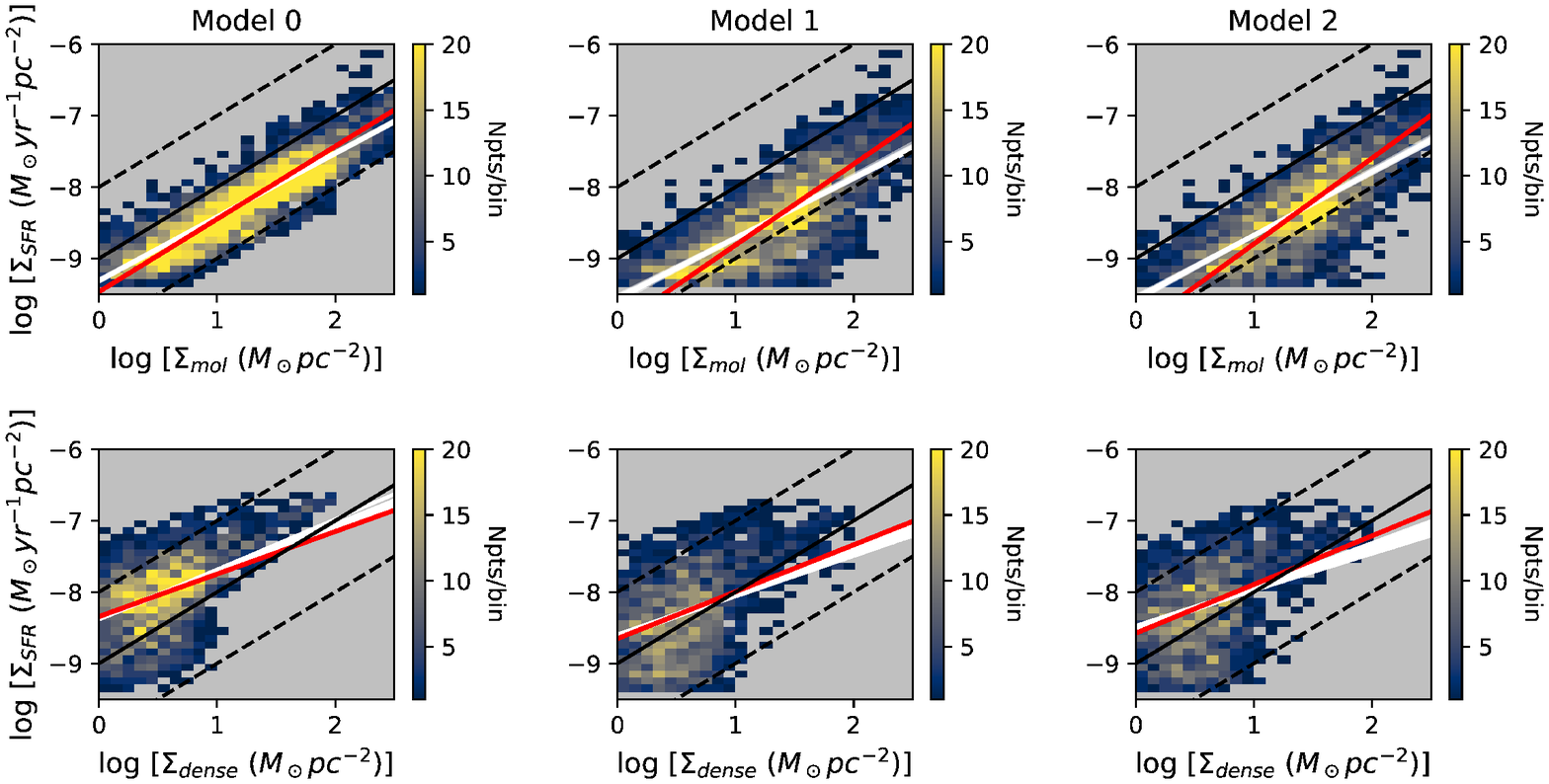}
\caption{(top row) Variation of $\Sigma_{SFR}$ with $\Sigma_{mol}$ for star formation rates calculated for (left) Model 0, 
(middle) Model 1 and (right) Model 2.  Images are 2 dimensional histograms of 
log($\Sigma_{SFR}$) and log($\Sigma_{mol}$).  
Solid red lines show the fit to all individual data points using bisector least squares method. 
The white solid lines show the model derived from 32 random draws of the slope and y-intercept posterior distributions. 
The black lines show the molecular gas 
depletion times from top to bottom of 10$^8$, 10$^9$, and 10$^{10}$ years.
(bottom row) 
 Two dimensional histogram showing the 
variation of $\Sigma_{SFR}$ with $\Sigma_{dense}$ for each SFR model.  
Red line represents the bisector fit to the 
data and the white lines show the Bayes' regression results from 32 random draws of the posterior distributions. 
\label{fig:fig5}}
\end{figure*}

M51 has been a primary target of studies exploring the resolved 
KS relationship between $\Sigma_{SFR}$ and $\Sigma_{mol}$ with varying results 
\citep{Kennicutt:2007, Bigiel:2008, Liu:2011, Chen:2015, 
Bigiel:2016, Leroy:2017}.  These differences arise in part due to angular resolutions and 
whether the analysis accounts for contributions from an older ($>$100~Myr) stellar population to the star formation 
tracer \citep{Liu:2011}.  In the top row of Figure~\ref{fig:fig5}, we 
show the KS relationship between $\Sigma_{SFR}$ and $\Sigma_{mol}={\alpha_{CO}}\int dv T(CO)$ M$_\odot$pc$^{-2}$
 derived for the three SFR models presented in \S3.2.  
Data with log($\Sigma_{SFR}$) less than -9.5 are excluded as the conversion to the 
star formation rate for these faint FUV and 24\micron\ luminosities is less reliable \citep{Leroy:2012}. 
We fit for the 
parameters $A$ and $N$ in the expression, log($\Sigma_{SFR}$)=A+N log($\Sigma_{mol}$) using 
bisector least squares in the python module {\tt bces} \citep{Nemmen:2012} 
that is based on \citet{Akritas:1996}.
We also calculate the posterior probability distribution for the set of regression parameters 
(denoted $A_B$, $N_B$)
using the {\tt emcee} package \citep{Foreman-Mackey:2013}.
Table~\ref{tab:table3} lists the best fit values for A and N, uncertainties, and scatter derived from the bisector fit 
  and the $\pm$2$\sigma$ range from the posterior distributions of A$_B$ and N$_B$. 
The bisector fits 
show the same trend identified by \citet{Liu:2011} in which the power law
index is near unity for SFR not corrected for contributions from the older stellar population as in Model 0.  For corrected 
SFR values (Model 1 and Model 2), the index steepens to 
 1.2.  
The Bayes' regression shows values of the slope posterior distributions to be linear or marginally sub-linear for all 3 SFR models.
\citet{Shetty:2013} applied hierarchical Bayesian regression on surface densities of star formation rates and 
molecular gas for a set of galaxies, including M51 for which they found a strongly sub-linear slope of 0.72 with a 
$\pm$2$\sigma$ range [0.62,0.83] over a narrower range of $\Sigma_{mol}$ and fewer points than the data used in 
this study.  

\begin{table*}[hbt]
\centering
\caption{Model Parameters log($\Sigma_{SFR}$)=A+Nlog($\Sigma_{X}$) \label{tab:table3}}
\begin{tabular}{ccccccc}
\hline
\hline
SFR Model & X & A & N & $\sigma$ & A$_B$ ($\pm$2$\sigma$) & N$_B$($\pm$2$\sigma$) \\
\hline
0 & mol & -9.48 (0.01) & 1.03 (0.01) & 0.28 & [-9.34, -9.30] & [0.88, 0.91] \\
1 & mol & -10.00 (0.02) & 1.16 (0.01) & 0.46 & [-9.66, -9.58] & [0.84, 0.89]\\
2 & mol & -10.06 (0.02) & 1.23 (0.01) & 0.46 & [-9.69, -9.60] & [0.90, 0.96] \\
0 & dense & -8.34 (0.01) & 0.60 (0.01) & 0.47 & [-8.39, -8.34] & [0.61, 0.68] \\
1 & dense & -8.65 (0.01) & 0.66 (0.01) & 0.54 & [-8.65, -8.58] & [0.54, 0.63] \\
2 & dense & -8.58 (0.01) & 0.68 (0.01) & 0.56 & [-8.54, -8.47] & [0.50, 0.60] \\
\hline
\end{tabular}
\end{table*}

\subsection{$\Sigma_{SFR}$ and $\Sigma_{dense}$}
The variation of the SFR with the dense gas mass is also an important relation as it 
probes the localized volumes of gas more spatially connected to star formation over the ensemble of clouds within 
the resolution element of the observations.  
The increased angular resolution 
of the LMT relative to other single dish telescopes provides sensitivity to the range of HCN luminosities 
5$\times$10$^4$ to 10$^6$ K km s$^{-1}$ pc$^2$, which is not well populated from previous studies. 
The LMT data complement 
the recent Northern Extended Millimeter Array (NOEMA) measurements that are also sensitive to this HCN luminosity range \citep{Querejeta:2019}. 

The bottom row of Figure~\ref{fig:fig5} shows the relationship between $\Sigma_{SFR}$ and  
$\Sigma_{dense}={\alpha_{HCN}}\int dv T(HCN)$ M$_\odot$pc$^{-2}$
for each SFR model in the form of a two dimensional histogram.  
Only points with $\Sigma_{SFR}$ greater than -9.5 are considered. 
We apply both bisector and Bayes' regression to 
the expression log($\Sigma_*$)=A+Nlog($\Sigma_{dense}$) and summarize the results in Table~\ref{tab:table3}. 
Sublinear relationships are derived for SFR Models 1 and 2 with slopes $\sim$0.65 (bisector) and 
a 2$\sigma$ posterior distributions range between 0.50 and 0.68 from the Bayes' regression.

\subsection{Star Formation Efficiency and $\Sigma_*$}
One of the key results to emerge from the EMPIRE survey is the anti-correlation of the star formation efficiency of dense gas, 
$SFE_{dense}=\Sigma_{SFR}/\Sigma_{dense}$, with 
the local environment properties: stellar mass surface density, molecular gas surface density, molecular gas fraction, and 
the  mid-plane pressure \citep{Gallagher:2018, Jimenez-Donaire:2019}.  
From 3 NOEMA fields of M51 with high angular resolution, 
\citet{Querejeta:2019} also identify an anti-correlation but with large scatter of points for a given $\Sigma_*$. 
The authors connect this trend to the Central Molecular Zone of the Milky Way in which there is a 
large reservoir of dense gas and a large dense gas fraction \citep{Jackson:1996, Jones:2012}
but a supressed star formation rate relative to the disk \citep{Longmore:2013, Kruijssen:2014,Barnes:2017, Lu:2019}.

To examine this dependence in our data, we apply spectral stacking within logarithmic bins of $\Sigma_*$ as in 
Figure~\ref{fig:fig4} 
to evaluate $\Sigma_{SFR}$, $\Sigma_{dense}$, and $\Sigma_{mol}$.  For a given log($\Sigma_*$)  bin, the 
star formation efficiencies 
are 
\begin{equation}
SFE_{dense} (Myr^{-1}) = 10^6 \frac{\Sigma_{SFR} (M_\odot pc^{-2} yr^{-1})}{\alpha_{HCN}\int dv <T(HCN)>)}
\end{equation}
and 
\begin{equation}
SFE_{mol} (Myr^{-1}) = 10^6 \frac{\Sigma_{SFR} (M_\odot pc^{-2} yr^{-1})}{\alpha_{CO}\int dv <T(CO)>)}, 
\end{equation}
where $\Sigma_{SFR}=\displaystyle\sum_{k=1}^{N_b} SFR_k/\sum_{k=1}^{N_b} A_k$, $A_k$ is the area per pixel in pc$^2$, 
$N_b$ is the number of pixels within the log($\Sigma_*$) bin, 
and $<T(HCN)>$ and $<T(CO)>$ are 
the stacked HCN and CO spectra respectively.  

\begin{figure*}
\centering
\includegraphics[width=0.75\textwidth]{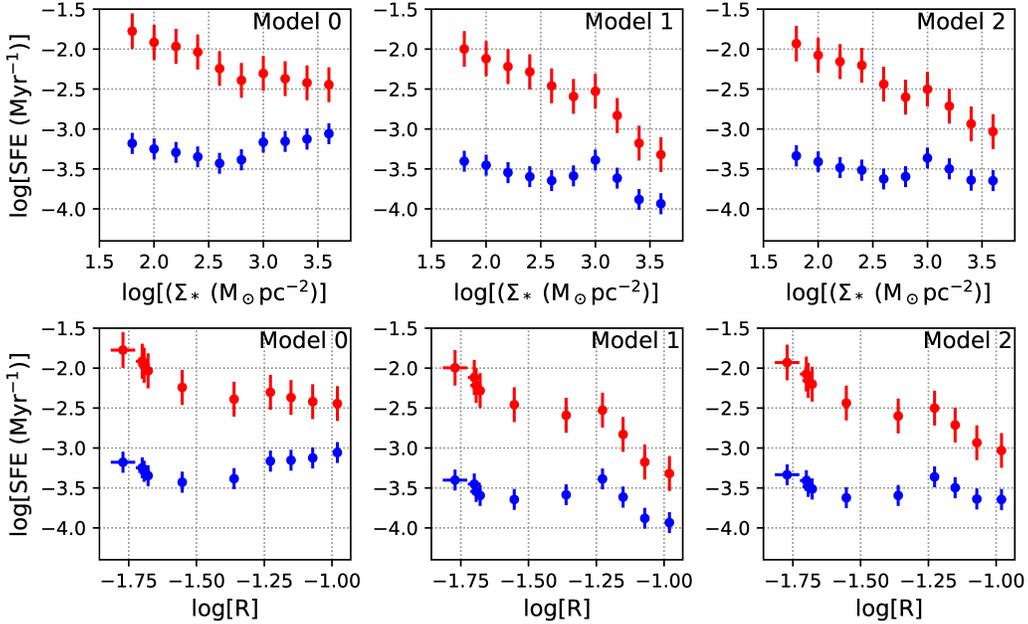}
\caption{(top row) Variation of star formation efficiencies for dense (red) and molecular (blue) gas within  
bins of stellar mass surface density for each SFR model described in \S3.2.  (bottom row) Star formation efficiency of dense and molecular gas with 
the ratio of HCN to CO luminosities derived in bins of stellar mass surface density.   Error bars mostly reflect the assumed uncertainties 
in $\alpha_{CO}$ and $\alpha_{HCN}$. 
\label{fig:fig6}}
\end{figure*}

Figure~\ref{fig:fig6} shows the variation of the 
dense gas (red points) and molecular (blue points) star formation efficiencies with 
log($\Sigma_*$) and the $R$ calculated within each log($\Sigma_*$) bin
for the 3 SFR models.  As found for many galaxies in the EMPIRE sample including M51, there is an overall  
decreasing relationship of $SFE_{dense}$ 
with increasing $\Sigma_*$ and $R$.  However, these profiles are not 
well described by a power law but rather, exhibit a broad inflection point 
over the log($\Sigma_*$) range 2.5 to 3.0. 
A similar inflection point is seen in the data presented by 
\citet{Gallagher:2018} for M51.   This inflection point is coincident with 
the stellar mass surface density at which $f_{DG}$ increases (see Figure~\ref{fig:fig4}).
$SFE_{mol}$ shows a mostly flat (Model 0) or slight decreasing (Models 1,2) profile 
 with increasing $\Sigma_*$ and $f_{DG}$ that is consistent with a linear or slightly sublinear 
relationship between $\Sigma_{SFR}$ and $\Sigma_{mol}$.  

\section{Discussion}
\subsection{Comparison to the Milky Way}
It is useful to compare these results on M51 with similar measures of $f_{DG}$ in the Milky Way where the improved spatial resolution enables
one to pinpoint the areas within a cloud responsible for the HCN emission. 
\citet{Jackson:1996} and \citet{Jones:2012} imaged the HCN and CO J=1-0 emission from the Central Molecular 
Zone (CMZ) of the Galaxy.  
The well known radio features, Sgr~A and Sgr~B, are bright in HCN emission but there is extended 
HCN emission throughout the CMZ that is distributed over a large velocity interval of $\pm$200~\kms. 
Averaged over the 500~pc aperture, comparable to the spatial resolution in this study, 
\citet{Jackson:1996} derive an HCN to 
CO ratio of 0.06 
corresponding to $f_{DG}$=0.14.   
Integrating over the entire CMZ, \citet{Jones:2012} measure a higher value of R=0.095, which scales to $f_{DG}$=0.22.
Figure~\ref{fig:fig2} shows HCN/CO=0.12, $f_{DG}$=0.28 in the nucleus of M51 and HCN/CO $>$0.056  
($f_{DG} >$=0.14) in the central bulge volume (see Figure~\ref{fig:fig3} and Figure~\ref{fig:fig4}).  

Observing HCN emission throughout the disk of the Milky Way is considerably more challenging given the required coverage and sensitivity. 
\citet{Helfer:1997} constructed an unbiased, undersampled survey of HCN and CO emission along the Galactic plane between 
longitudes 15.5$^\circ$ and 55.5$^\circ$, spaced by 1$^\circ$ intervals and b=0$^\circ$  with the NRAO 12~m telescope.  
Their HCN to CO ratio shows a flat profile with Galactic radius with an 
average value of 0.026 ($f_{DG}$=0.06).  From this low value of R with respect to HCN/CO values observed towards resolved, active star forming regions 
(0.1-0.3) within Galactic molecular clouds, they inferred that the HCN emission was originating from the more extended, 
lower density (10$^3$~\cc) components 
of the clouds.  
\citet{Evans:2020} imaged HCN J=1-0 emission from 6 molecular clouds in the first quadrant of the Milky Way.  
They determined HCN to $^{13}$CO ratios between 0.04 and 0.12 
but only a fraction (0 to 57\%) of the HCN luminosity comes from regions with $A_v > 8$ that 
is assumed to reflect high gas volume density where stars form.  
\citet{Battisti:2014} estimated the dense gas mass fraction for 344 clouds in the first quadrant of 
the Milky Way using ground-based thermal 1.1mm dust continuum emission as a measure of dense gas mass 
and $^{13}$CO J=1-0 emission for the total cloud mass. 
While optically thin dust emission is a direct tracer of gas column density, data processing methods used for ground-based 
millimeter continuum measurements to account for 
atmospheric contributions also 
remove extended emission from the larger clouds
 depending on the size of the cloud relative to the field of view of the bolometer 
array \citep{Ginsburg:2013}.  The resultant image identifies regions of overdensity 
within a cloud that typically correspond to dense clumps and filaments.
\citet{Battisti:2014} found $f_{DG}$ values of 0.11 for all dust emission and 0.07 for subregions with mass surface densities  
greater than 200 M$_\odot$pc$^{-2}$ that likely represents higher volume densities. 
The fractions are found to be independent of cloud mass and mass surface density. 

Figure~\ref{fig:fig3} and Figure~\ref{fig:fig4} show $f_{DG}$ values within the disk of M51 with a mean value of 0.05. This mean value is comparable 
to Milky Way values found by \citet{Helfer:1997} and inferred from \citet{Evans:2020} that point to a 
lower density ($\sim$10$^3$~\cc) origin 
where the HCN rotational energy levels are subthermally excited resulting in weaker emission.   The density of this component 
is still larger than the mean density of molecular clouds that is inferred from \co\ and \coa\ emission.  

\subsection{$f_{DG}$, Spiral Structure and Stellar Bulge}
A surprising result from our stacking analysis is the near-constant value of $f_{DG}\sim0.05$ 
between spiral arm and interarm regions and throughout the disk component of M51. 
The actions of a spiral density wave are gas compression and 
increased interstellar turbulence \citep{Kim:2002} that contribute to the development of dense regions 
within 
molecular clouds.  
So one might expect to find elevated values of $f_{DG}$ in 
the spiral arms relative to those derived in the interarm regions. 

The expected higher $f_{DG}$ values in spiral arms can be resolved with our data 
if the conversion factor 
between CO luminosity and \htwo\ mass is different in these respective domains.
There is evidence that the CO conversion factor is $\sim$2 times larger 
in the interarm regions than spiral arms 
in M51 and M83 \citep{Wall:2016}. 
Adopting this arm-interarm dependence of the conversion factor 
would lead to $f_{DG}$ values 2 times higher in the spiral arms than 
interarm regions. 

Alternatively, the constant dense gas mass fraction throughout the disk could result from 
the density thresholds 
required to excite the HCN and CO lines within a population of stratified, self-gravitating clouds
\citep{Elmegreen:2018}.   For each spectral line, there is a unique cloud radius, $r_c$,
 at which the volume 
number density
is equal to the line's effective critical density, $n_c$.  For a 1/$r^2$ density profile, the 
mass within this radius is proportional to $r_c$.  Since $r_c$ scales with $n_c^{-1/2}$, 
then the 
dense gas fraction derived from HCN and CO measurements is 
$(n_{CO}/n_{HCN})^{1/2}$, 
where $n_{CO}$ and $n_{HCN}$ are the critical densities for CO and HCN respectively. 
For critical densities of 100 \cc\ for CO and 10$^4$ \cc\ for HCN, $f_{DG}=0.1$ which is a reasonable 
estimate for the M51 values in the disk given the approximations of the effective critical densities
described in \S4. 

\begin{figure*}[htb!]
\centering
\includegraphics[width=0.75\textwidth]{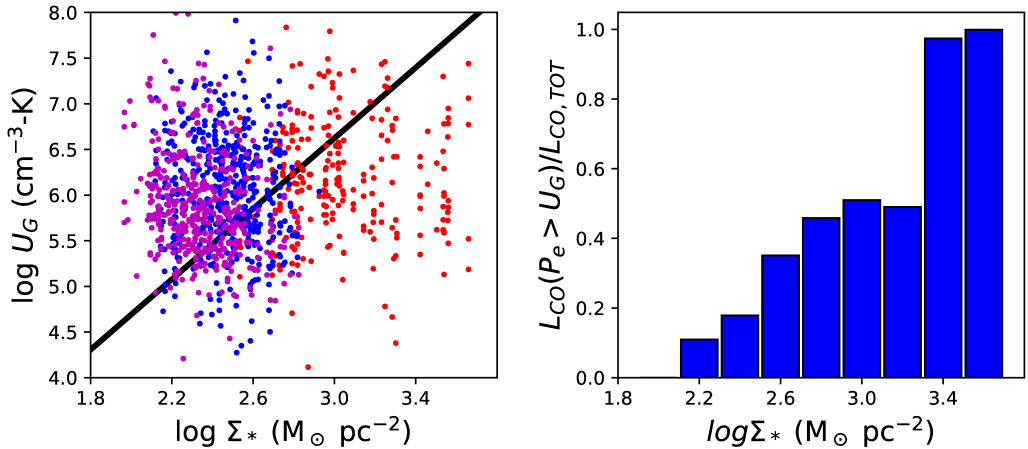}
\caption{(left) 
The self-gravitational energy densities for giant molecular clouds and cloud complexes in M51 within 
the central bulge region (red)
spiral arms(blue), and interarm regions (magenta) from the compilation of molecular clouds in M51 by \citet{Colombo:2014}.  The solid line shows 
the profile of kinematic pressure from the weight of the stars and molecular gas at the boundary of a cloud.
Points below this line represent diffuse clouds.
(right) The fraction of CO luminosity from diffuse clouds for each log $\Sigma_*$ bin. 
\label{fig:fig7}}
\end{figure*}

Figure~\ref{fig:fig4} shows the dense gas fraction begins to increase 
for log($\Sigma_*) > 2.5$.  In this domain, the pressure, $P_e$, from the weight of stellar and gas components 
may exceed the gravitational energy density of a cloud, $U_G=\frac{\pi G}{2}\Sigma_{cloud}^2$, where $\Sigma_{cloud}$ 
is the molecular gas surface density of an individual cloud.  Clouds that satisfy the condition $P_e > U_G$ are labeled as diffuse clouds \citep{Elmegreen:2018}. 
In Figure~\ref{fig:fig7} we 
show the variation of $U_G$ with log($\Sigma_*$) 
for giant molecular clouds and cloud complexes in M51 residing in the central region, 
spiral arms and interarm regions 
from the catalog constructed by \citet{Colombo:2014}.  
The $\Sigma_*$ value for each cloud is taken from the stellar mass surface density at the position of the cloud.
The total mid-plane pressure can be approximated as 
\begin{equation}
P_{tot}=\frac{\pi}{2}G\Sigma_{gas}\Bigl(\Sigma_{gas}+\Sigma_*\frac{c_{g,tot}}{c_s}\Bigr)
\label{eq:pressure}
\end{equation}
where $c_s$ and $c_{g,tot}$ are the star and gas velocity dispersions \citep{Elmegreen:1989}. 
At the cloud boundary, the kinematic pressure, $P_e=P_{tot}/(1+\alpha_\circ+\beta_\circ)$ where 
$\alpha_\circ=0.46$ and $\beta_\circ=0.30$ are the relative cosmic ray and magnetic field energy densities \citep{Draine:2010}.
The measured gas velocity dispersion also requires a correction for cosmic rays and the interstellar magnetic field such that 
$c_{g,tot}=c_g/(1+\alpha_\circ+\beta_\circ)^{1/2}$.
From averages of $\Sigma_{mol}$ and $\Sigma_*$ over 1~kpc wide annular radii, we derive the relationship 
$\Sigma_{mol}=0.35\Sigma_*^{0.95}$.  
Assuming fully molecular gas ($\Sigma_{gas}=\Sigma_{mol}$), 
$c_s$=20~\kms\ and $c_g$=10~\kms, we construct the profile of pressure with $\Sigma_*$ from equation~\ref{eq:pressure}.
This 
profile of pressure with log($\Sigma_*$) is shown as the 
solid black line in Figure~\ref{fig:fig7}. 
For low values of log($\Sigma_*$) corresponding to the M51 disk (magenta and 
blue points), most 
clouds are above the pressure line consistent with self-gravitating objects. 
In the central bulge, most of the defined objects (red points) 
lie below the pressure 
profile and are considered as diffuse clouds. 

From this information, the fraction of CO luminosity contributed from 
clouds with $P_e > U_G$ for each $\Sigma_*$ bin are calculated (see right panel of Figure~\ref{fig:fig7}).
For bins of stellar mass surface densities greater than 500~\msun~pc$^{-2}$, $>$45\% of the molecular cloud mass resides within  diffuse clouds.
These fractions are also derived for the set of clouds in each region described by \citet{Colombo:2014}. 
The mass fraction of diffuse clouds is 91\%  in the central region, 14\% in spiral arms, and 32\% in the interarm regions.

The increasing fraction of molecular cloud mass in the diffuse state can account for the rise of the dense gas mass fraction at high stellar 
mass surface densities.  
In the central 2~kpc, the neutral gas is already 
fully molecular \citep{Jimenez-Donaire:2019} so there is no available atomic hydrogen to compress and contribute 
to the molecular mass.
The external pressure can 
act upon the molecular gas to increase the mean density of the now diffuse molecular 
clouds.  This compression  
leads to an increase in the dense gas mass fraction that is seen in Figure~\ref{fig:fig3} and Figure~\ref{fig:fig4}.  
At even higher stellar mass surface densities within the central bulge, the pressure may even exceed 
the self-gravity energy density 
of the dense gas 
component.  We note that for such conditions, our adopted value of $\alpha_{HCN}$ is an 
upper limit \citep{Gao:2004a}, which may increase $SFE_{dense}$ values in the highest stellar mass surface 
density bins and flatten the $SFE_{dense}$ profile.

\subsection{Environment and Star Formation Efficiency}
The profiles of $f_{DG}$ and the star formation efficiency with increasing stellar mass surface density are clues for 
understanding the role of galactic environment on the dynamical state of molecular clouds and the 
impact on the formation of new stars in galaxies.  
In M51, we find higher star formation 
efficiencies of dense gas in the disk, where the stellar mass surface density and mid-plane 
pressure are low, and smaller values of $SFE_{dense}$ in the central bulge, where $\Sigma_*$ and 
pressures are high.  
Similar 
results have been established by earlier studies of 
M51 and other galaxies 
\citep{Bigiel:2016, Gallagher:2018, Jimenez-Donaire:2019}. 
The transition between these regimes occurs between log($\Sigma_*)\sim$2.5-3.0, where an increasingly larger fraction of the molecular mass 
resides within the diffuse cloud population 
(Figure~\ref{fig:fig7}).  

\citet{Gallagher:2018} and 
\citet{Jimenez-Donaire:2019} propose that the lower dense gas star formation efficiencies 
in the high pressure environment are a 
consequence of a higher mean density of clouds such that star formation occurs in 
regions of even higher volume densities than is traced by the HCN J=1-0 line. The smaller volume and mass fraction of such overdense regions 
implies a larger amount of lower density material that is contributing to the HCN J=1-0 luminosity but is not actively forming stars, 
which leads to lower values of $SFE_{dense}$.  

The proposed shift of star formation activity to higher density regions can be understood if one 
considers self-gravity of the dense protocluster clumps and protostellar cores as a fundamental requirement for the 
production of new stars.  The formation of stars occurs when such structures collapse under their own self-gravity rather than 
implode from an external pressure disturbance. 
For high pressure environments like the central bulge of M51, the condition 
$U_G > P_e$ is realized at much 
higher surface and volume densities than are probed by the HCN J=1-0 line. 

A simple test for this description is to measure the variation of the star 
formation efficiency in dense gas using a tracer with a much higher critical density than 
the HCN J=1-0 
transition used in this and other studies.  These data would directly trace the higher volume density and higher surface density 
regions that remain 
self-gravitating even under large external pressures.  With a tracer of very high densities 
such that $U_G > P_e$, 
one would expect to see the $SFE_{dense}$ profile to 
be flat or rising 
with increasing $\Sigma_*$ and pressure.

\section{Conclusions}
Using 
\co\ and HCN J=1-0 data with 582~pc resolution collected by the Large Millimeter Telescope and ancillary data from {\it GALEX}, {\it Spitzer}, SDSS, and 2MASS, we have 
investigated the variation of the dense gas mass fraction within different environments of the M51 galaxy and the impact on star formation.   
Within the disk component, $f_{DG}$ can vary along a spiral arm segment but on 
average, dense gas fractions are comparable in spiral 
arms and interarm regions. 
The dense gas mass fraction rises steeply in the central 
bulge, where the stellar mass surface density is greater than 500 \msun/pc$^{-2}$. 
The star formation efficiency of dense gas decreases with increasing stellar mass surface surface density 
with an inflection point over $\Sigma_* \sim$ 400-1000 \msun pc$^{-2}$ that may mark a transition from self-gravitating 
structures to diffuse clouds and clumps.  
At the highest pressures in the central bulge,  star formation is limited to more compact, high density, self-gravitating structures that 
do not significantly contribute to the HCN J=1-0 luminosity, which may lead to the measured decrease in the star formation efficiency of dense gas.

\acknowledgments
The authors thank the referee Frank Bigiel who provided valuable comments that improved the manuscript. 
Authors MH, BG, and DC acknowledge support from NSF grant AST-1907791.
A.~A. acknowledges the support of the Swedish Research Council, Vetenskapsr\aa{}det, 
and the Swedish National Space Agency (SNSA).
This publication makes use of data products from the Two Micron All Sky Survey, which is a joint project of the University of Massachusetts and the Infrared Processing and Analysis Center/California Institute of Technology, funded by the National Aeronautics and Space Administration and the National Science Foundation.
NASA/IPAC Extragalactic Database (NED) is funded by the National
Aeronautics and Space Administration and operated by the California Institute of
Technology.   

\vspace{5mm}
\facilities{LMT(SEQUOIA)}

\software{astropy \citep{astropy:2013}, emcee \citep{Foreman-Mackey:2013}, bces \citep{Nemmen:2012}}
\clearpage
\appendix
\section{Galfit Model Parameters}{\label{appendixA}
To estimate the contributions from the old stellar population, we decompose the {\it Spitzer} 3.6\micron\ image of surface 
brightness into model components using {\tt Galfit} \citep{Peng:2010}. The parameters for each model component are 
described in \citet{Peng:2010}.  The values of the best-fit  
parameters derived for M51 for SFR Model 1 and SFR Model 2 are listed in 
Table~\ref{tab:table4} and Table~\ref{tab:table5}. 
Figure~\ref{fig:fig8} shows the superpositions of all model components that are fit to the {\it Spitzer} 3.6\micron\ 
intensities and from which the cirrus contributions are estimated for SFR Model~1 and Model~2.  
\begin{figure*}
\centering
\includegraphics[width=0.75\textwidth]{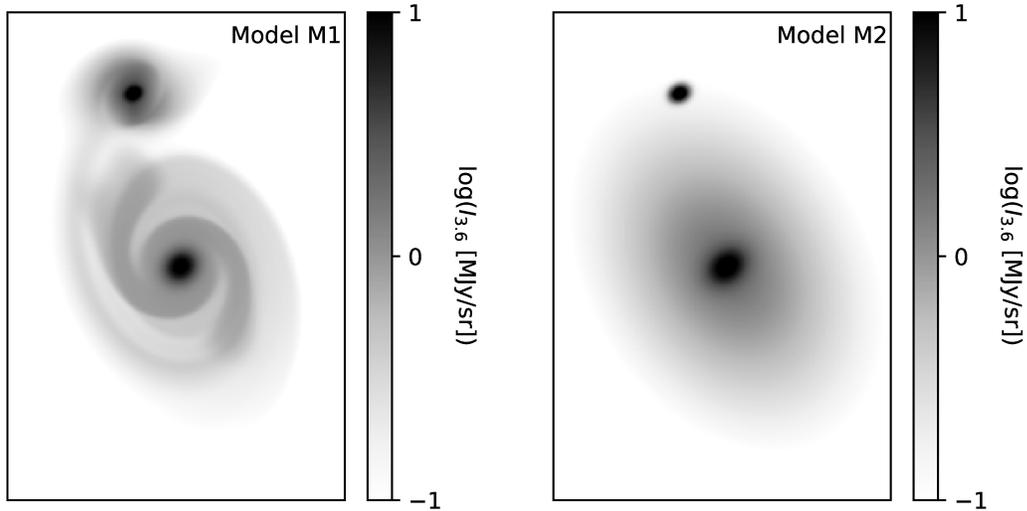}
\caption{{\tt Galfit} models of 3.6\micron\ surface brightness for (left) Model 1 with bulge and spiral arm components and (right) Model 2 with 
bulge component and an exponential disk with no spiral arms. 
\label{fig:fig8}}
\end{figure*}

\begin{table}
\centering
\caption{Galfit parameters for Model 1\label{tab:table4}}
\begin{tabular}{lccccccccccc}
 \hline
 \hline
Parameter & \# & sersic & ${\delta}$x('') & ${\delta}$y('') & mag & $r_e(')$ & n & q & $\theta_{P.A.}$ (deg) & & Comment\\
          &   & power & ... & $r_{in}$(') & $r_{out}$(') & $\theta_{rot}$ (deg) & $\alpha$ & $\theta_{incl}$(deg) & $\theta_{sky}$ (deg) \\
          &  & fourier & a$_1$ & $\phi_1$ (deg) & a$_3$ & $\phi_3$ (deg) & a$_4$ & $\phi_4$ (deg) & a$_5$ & $\phi_5$ (deg) \\
 \hline
M51a & 1 & sersic & -0.09 & -0.18 & 13.0 & 0.040 & 1.18 & 0 & -65.5  & & bulge\\
     & 2 & sersic & 0.17 & 0.31 & 10.3 & 0.260 & 0.67 & 0 & -47 & &  bulge\\
     & 3 & sersic & 3.4  & 2.0 & 8.9 & 2.780 & 0.35 & 0 & -26  & & outer spiral\\
     &   & power &  & -1.29 & 4.28 & -791 & 0.29 & 40 & -74 &  \\
     &   & fourier & -0.117 & 55.3 & -0.006 & -18.9& 0.041 & -33.7 & 0.023 & 28.2 \\
     & 4 & sersic & -3.3 & 0.8 & 10.0 & 1.88 & 0.14 & 0 & 19 & & inner spiral\\
     &   & power &  & 0.66 & 2.34 & -194 & -0.11 & -0.01 & 29 &  \\
     &   & fourier & -0.122 & 11.4 & -0.056 & -55.5 & 0.058 & 3.9 & 0.014 & 4.1 \\
M51b & 5 & sersic & -69.7 & 255.4 & 11.9 & 0.0428 & 0.38 & 0 & -82&  &bulge  \\
     & 6 & sersic & -69.4 & 255.5 & 11.0 & 0.1107 & 0.74 & 0 & -64  & & bulge \\
     & 7 & sersic & -72.1 & 252.4 & 9.9 & 0.900 & 0.72 & 0 & -25  & &  spiral \\
     &   & power & & 0.880 & 1.08 & 32 & 1.6 & 42 & 49 &  \\
     &   & fourier & -0.128 & 113.4 & -0.070 & -1.0 & 0.020 & 18.6 & 0.018 & 8.3 \\
 \hline
\end{tabular}
\end{table}

\begin{table}
\centering
\caption{Galfit parameters for Model 2\label{tab:table5}}
\begin{tabular}{lccccccccccc}
 \hline
 \hline
Parameter & \# & sersic & ${\delta}$x('') & ${\delta}$y('') & mag & $r_e(')$ & n & q & $\theta_{P.A.}$ (deg) & & Comment\\
          &   & power & ... & $r_{in}$(') & $r_{out}$(') & $\theta_{rot}$ (deg) & $\alpha$ & $\theta_{incl}$(deg) & $\theta_{sky}$ (deg) \\
          &  & fourier & a$_1$ & $\phi_1$ (deg) & a$_3$ & $\phi_3$ (deg) & a$_4$ & $\phi_4$ (deg) & a$_5$ & $\phi_5$ (deg) \\
 \hline
M51a & 1 & sersic & -0.08 & -0.17 & 13.1 & 0.0340 & 0.92 & 0 & -82  & &bulge\\
     & 2 & sersic & 0.17 & 0.29 & 9.9 & 0.3200 & 0.76 & 0 & -51 & & bulge\\
     & 3 & sersic & 0.17  & 0.29 & 8.0 & 2.4 & 0.82 & 0 & 31  &  & disk\\
M51b & 4 & sersic &-69.6  & 255.4 & 11.6 & 0.04 & 0.38 & 0 & -81 & &bulge\\
     & 5 & sersic & -69.5 & 255.4 & 10.9 & 0.113 & 0.75 & 0 & -63  & & bulge \\
     & 6 & sersic & -68.7 & 259.3 & 9.6 & 0.96 &  0.70 & 0 & -26  & & spiral \\
     &   & power & & 0.927 & 1.38 & 39 & 1.7 & 50 & 51 &  \\
     &   & fourier & 0.060 & 96.3 & 0.023 & -57.0 & 0.056 & 20.5 & 0.016 & 10.6 \\
 \hline
\end{tabular}
\end{table}

\bibliography{m51.bib}{}
\bibliographystyle{aasjournal}

\end{document}